\documentclass[12pt]{article}
\usepackage{graphics}

\title{
Geometrical and physical properties of maths-type $q$-deformed coherent states
for $0<q<1$ or $q > 1$}

\author{C.\ Quesne $^{a,}$\thanks{Corresponding author.
{\sl E-mail addresses}: penson@lptl.jussieu.fr (K.A.\ Penson), cquesne@ulb.ac.be (C.\
Quesne), tkachuk@ktf.franko.lviv.ua (V.M.\ Tkachuk)}\ , K.A.\ Penson $^b$,  V.M.\ 
Tkachuk $^c$\\
{\small\sl $^a$ Physique Nucl\'eaire Th\'eorique et Physique Math\'ematique, 
Universit\'e Libre de Bruxelles,} \\ 
{\small \sl Campus de la Plaine CP229, Boulevard~du Triomphe, B-1050
Brussels, Belgium}\\
{\small\sl $^b$ Universit\'e Pierre et Marie Curie, Laboratoire de Physique Th\'eorique des
Liquides, }\\
{\small\sl CNRS UMR 7600, Tour 16, 5i\`eme \'etage, 4, place Jussieu, F-75252 Paris
Cedex 05, France}\\ {\small\sl $^c$ Ivan Franko Lviv National University, Chair of
Theoretical Physics,}\\ {\small\sl 12, Drahomanov Street, Lviv UA-79005, Ukraine}}
\date{ }
\begin{document}
\baselineskip=22pt plus 1pt minus 1pt
%%%%%%%%%%%%%%%%%%%%%%%%%%%%%%%%%%%%%%%%%%%%%%%%%%%%%%%%%%
\maketitle


\par\bigskip
%
%========================================================================
%
We wish to compare the geometrical and physical properties of the
maths-type coherent states for $q>1$~\cite{cq} with those of the same for
$0 < q < 1$. Many mathematical properties of the latter were indeed studied by Arik and
Coon~\cite{arik}, but as far as we know, their remaining properties  have not been
considered so far in the literature and will therefore be reviewed here.\par
%
%-----------------------------------------------------------------------------------
%
All the formulas presented in Sect.~3 of Ref.~\cite{cq} are valid whatever range of $q$
values we deal with. The only thing to be remembered when applying them to $q$ values
in the interval $(0,1)$ is that the $z$ values are then restricted to the disc of radius
$[\infty]^{1/2} = (1-q)^{-1/2}$.\par
%
%----------------------------------------------------------------------------------
%
The metric factor $\omega_q(t) = d \langle N\rangle_q/dt$ (where $t \equiv |z|^2$)
behaves as $\omega_q(t) \simeq 1 + 2(1-q) (1+q)^{-1} t + O(t^2)$ for small $t$
values. It is therefore larger than $\omega(t) = 1$, corresponding to the flat geometry of
the conventional CS. Numerical calculations confirm this result for higher $t$ values:
$\omega_q(t)$ turns out to be an increasing function of $t$ and of $1-q$. The behaviour
of $\omega_q(t)$ as a function of $t$ is therefore opposite in the two ranges $0 < q < 1$
and $q > 1$.\par
%
%--------------------------------------------------------------------------------------------------------
%
The same conclusion applies to the Mandel parameter $Q_q = [(\Delta N)_q^2 - \langle N
\rangle_q]/ \langle N\rangle_q$. It indeed behaves as $Q_q(t) \simeq (1-q)t/(1+q) +
O(t^2)$ for small $t$ values. As shown in Fig.~1, it remains positive for
higher $t$ values, so that the photon number distribution is super-Poissonian instead
of being sub-Poissonian as for $q > 1$.\par
%
%------------------------------------------------------------------------------------------------------
%
The discussion gets somewhat more involved for the variance $(\Delta X)_q^2$ of the
Hermitian quadrature operator $X$, given by Eq.~(26) in Ref.~\cite{cq}. The
coefficient of $2 ({\rm Re}\, z)^2$ in this equation, behaving as $\sqrt{2/(1+q)} -1
+ O(t)$ for small $t$ values, can be checked to be positive over a wide range. As a
consequence, for a given $t$, the maximum squeezing in $X$ can be achieved when $z$ is
imaginary (vs.~$z$ real for $q > 1$). On setting $z = {\rm i} \sqrt{t}$, we find that for
small $t$ values, $R_q(t) = 2 (\Delta X)_q^2 \simeq 1 + 2\left[1 -
\sqrt{2/(1+q)}\right]t + O(t^2)$ \footnote{There is an unfortunate typo in the
corresponding result for $q > 1$ and $z = \sqrt{t}$, given in Ref.~\cite{cq}. The correct
formula reads $R_q(t)  \simeq 1 - 2\left[1 - \sqrt{2/(1+q)}\right]t + O(t^2)$.}, showing
the presence of squeezing as in the $q > 1$ case. This is confirmed in Fig.~2, where
$R_q(t)$ is plotted against $t$ for $z = {\rm i} \sqrt{t}$ and several $q$ values.\par
%
%-------------------------------------------------------------------------------------------
%
The behaviour of the signal-to-quantum noise ratio $\sigma_q = \langle X
\rangle_q^2/(\Delta X)_q^2$ is still more complicated because for a given $t$ value,
the numerator $\langle X \rangle_q^2$ is maximum for real $z$, while the denominator
$(\Delta X)_q^2$ is minimum for imaginary $z$. We have therefore to consider
nonvanishing values of both ${\rm Re}\, z$ and ${\rm Im}\, z$. A three-dimensional plot
of $\sigma_q$ shows that the latter increases with both ${\rm Re}\, z$ and ${\rm Im}\,
z$, but that it always remains smaller than $4 \langle N \rangle_q$, corresponding
to the value $4N_s$ attained for the conventional CS. There is therefore no improvement
on the conventional CS value, contrary to what happens for $q > 1$.\par
%
%-------------------------------------------------------------------------
%
Let us finally consider the deformed commutation relation (28) in Sect.~4 of~\cite{cq}
and assume this time that $\alpha$ and $\beta$ are both negative. It is then easy to see
that Eq.~(29) is replaced by $q = (1 - \hbar \sqrt{|\alpha| |\beta|}) / (1 + \hbar
\sqrt{|\alpha| |\beta|})$, where $q$ varies in the interval (0, 1). For $|\alpha| = m^2
\omega^2 |\beta|$, the harmonic oscillator Hamiltonian (31) becomes equivalent to the
Arik and Coon $q$-deformed oscillator Hamiltonian of frequency $\frac{1}{2} (1+q)
\omega$, so that the CS associated with the latter are intelligent CS for the former.
Such CS, however, have no application in connection with problems in string
theory and quantum gravity because for negative $\alpha$ and $\beta$,
Eq.~(28) of~\cite{cq} does not lead to nonzero minimal uncertainties in position
and momentum. This of course contrasts with the case $\alpha, \beta >0$ or
$q>1$, described in~\cite{cq}.
%
%====================================================
%
\section*{Acknowledgments}

C.Q.\ would like to thank J.R.\ Klauder for an interesting discussion. She is a Research
Director of the National Fund for Scientific Research (FNRS), Belgium.\par
%
%============================================================
%

%
%========================================================================
% 
\newpage
\section*{Figure captions}

{}Fig.\ 1. The Mandel parameter $Q_q(t)$ as a function of $t = |z|^2$ for $q = 0.98$
(solid line), $q = 0.96$ (dashed line) and $q = 0.94$ (dot-dashed line).

\noindent
{}Fig.\ 3. The variance ratio $R_q(t)$ as a function of $t = |z|^2$ for imaginary $z$
and $q = 0.98$ (solid line), $q = 0.96$ (dashed line) and $q = 0.94$ (dot-dashed line).
\par
%
%======================================================================
%
%\newpage
%\vspace*{15cm}
%\centerline{Figure 1}
%
%------------------------------------------------------------------------------------------------------
%
%\newpage
%\vspace*{15cm}
%\centerline{Figure 2}
%------------------------------------------------------------------------------------------------------
\newpage
\begin{picture}(160,100)
\put(35,0){\mbox{\scalebox{1.0}{\includegraphics{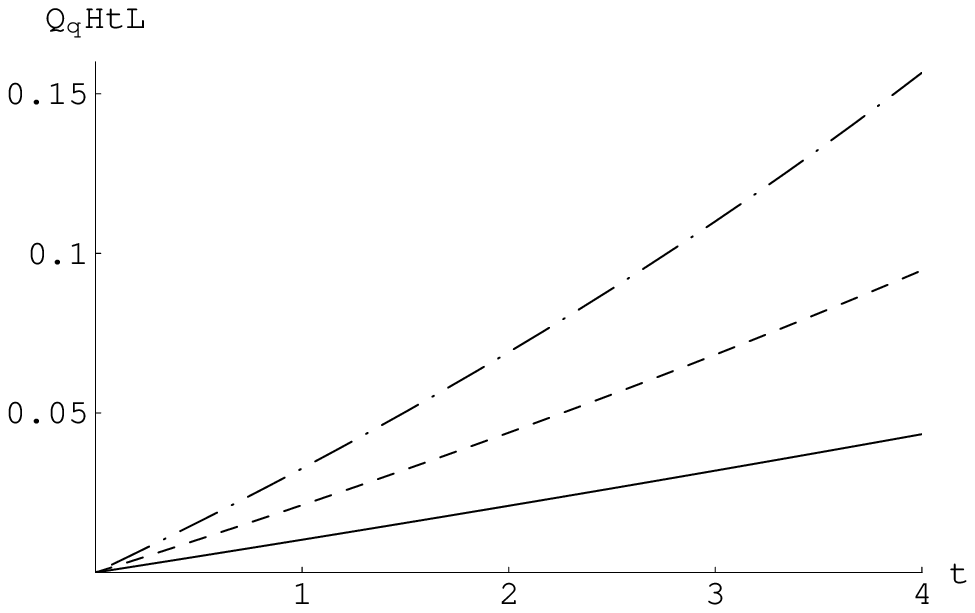}}}}
\end{picture}
%\vspace*{5cm}
%\centerline{Figure 1}
%
%----------------------------------------------------------------------
%  
\newpage
\begin{picture}(160,100)
\put(35,0){\mbox{\scalebox{1.0}{\includegraphics{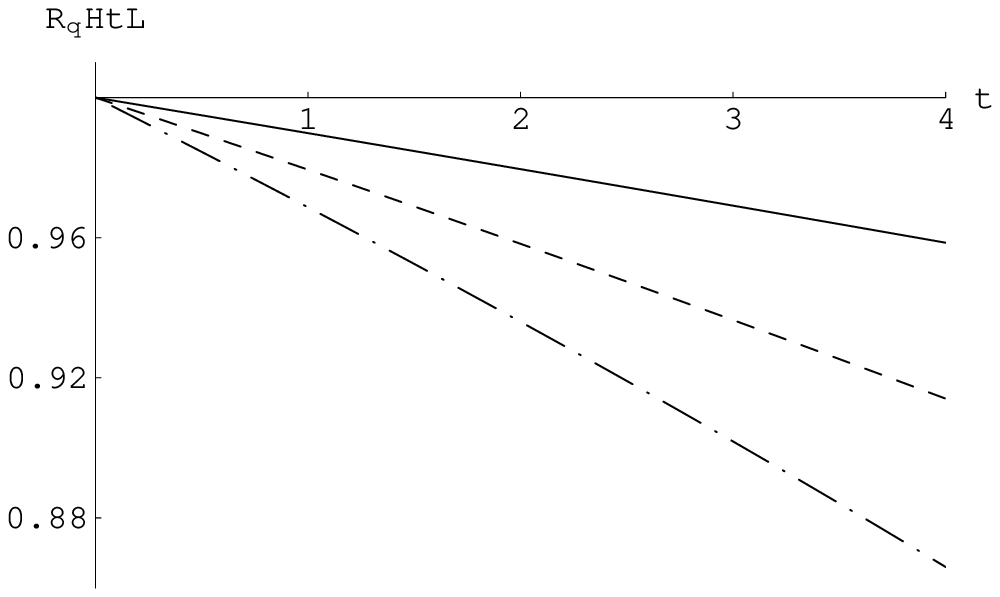}}}}
\end{picture}
%\vspace*{5cm}
%\centerline{Figure 2}
%
%----------------------------------------------------------------------
% 
 
\end{document}